\documentclass[journal=jacsat,manuscript=article]{achemso}
\usepackage[version=3]{mhchem} 
\usepackage{graphicx}

\author{Joshua T. Berryman}
\email{josh.berryman@uni.lu}
\author{Tanja Schilling}
\affiliation[University of Luxembourg]{University of Luxembourg, Luxembourg}

\title{Free Energies by Thermodynamic Integration Relative to an Exact Solution, Used to Find the Handedness-Switching Salt Concentration for DNA}


\begin{document}
\begin{abstract}
Sets of free energy differences are useful for finding the equilibria of chemical reactions, while absolute free energies have little physical meaning. However finding the relative free energy between two macrostates by subtraction of their absolute free energies is a valuable strategy in certain important cases.  We present calculations of absolute free energies of biomolecules, using a combination of the well-known Einstein Molecule method (for treating the solute) with a conceptually related method of recent genesis for computing free energies of liquids (to treat the solvent and counterions).  The approach is based on thermodynamic integration from a detailed atomistic model to one which is simplified but analytically solvable, thereby giving the absolute free energy as that of the tractable model plus a correction term found numerically.  An example calculation giving the free energy with respect to salt concentration for the B- and Z-isomers of all-atom duplex DNA in explicit solvent and counterions is presented.  The coexistence salt concentration is found with unprecedented accuracy.
\end{abstract}

\section{Introduction}

{\bf Overview:} Recent improvements of atomistic forcefields for nucleic acids \cite{Perez2007} and advances in a particular family of thermodynamic integration techniques \cite{Eike2006,Tyka2007,Cecchini2009,Schilling2009} together encourage a return to what we believe should be considered, due to its combination of simplicity and intractability, as a canonical model system to benchmark computational methods for biomolecules and polyelectrolytes: the handedness-switching B to Z isomerisation of DNA. 

Here we present a new addition to the progression of free energy methods beginning from the ``Einstein Crystal'' of Frenkel \& Ladd \cite{Frenkel1984}, followed by the ``Einstein Molecule'' \cite{Vega2007}, the ``Confinement Method'' \cite{Tyka2007} and a recently introduced analytical reference-liquid model \cite{Schilling2009,Schmid2010}. We demonstrate this new method using the example of the transition between B and Z DNA.  In the absence of an existing umbrella term for this philosophy of free energy calculation, it is referred to here as Thermodynamic Integration relative to an Exact Solution (TIES) because it relies on performing a Thermodynamic Integration (TI) calculation  with a realistic (but intractable) model at one endpoint of the integration path and an exactly solvable (ES) model at the other.  The principal technical advance within the field of TIES to be presented here is in the adaptation of a recently developed reference liquid model \cite{Schilling2009,Schmid2010} such that it can be used for triangular water in a molecular dynamics simulation, and successful combination of this adapted liquid model with the existing Einstein Molecule (EM) model for the solute. 

For the `realistic', but intractable, end of the integration path, the AMBER simulation code \cite{Case2010} and selected AMBER forcefield parameters were used.  The details of this quantitative all-atom treatment (with Coulomb, bonded and van der Waals interactions) are given in the methods.

 The results of the example calculation are found to be of groundbreaking accuracy for the example considered, although they were computationally expensive and scope for future gains in efficiency is acknowledged.

\begin{figure}
\includegraphics[width=0.5\textwidth]{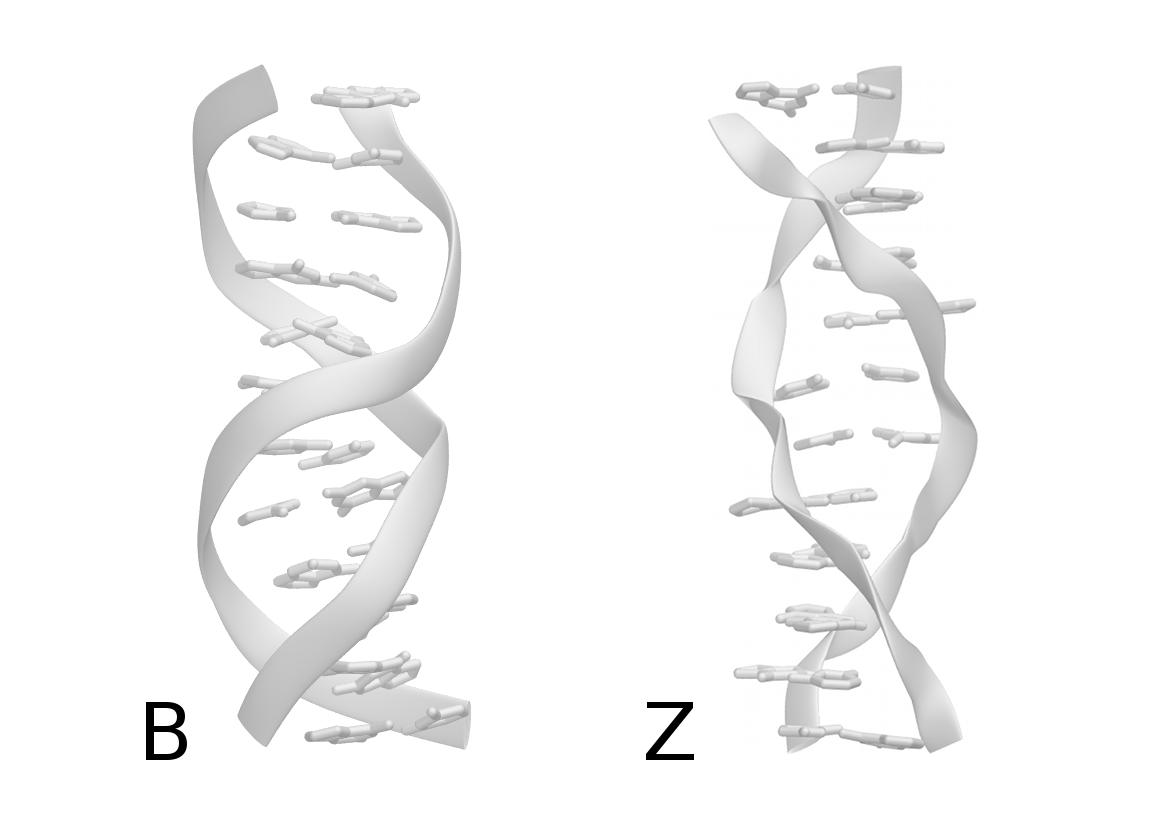}
\caption{B and Z isomers of [d(CG)$_6$]$_2$ DNA, after 12ns equilibration in 1 Molar NaCl. The backbone phosphate groups are closer together in the more compact Z isomer, and the base-stacks have greater solvent exposure.}
\label{fig:structs}
\end{figure}

{\bf B-Z DNA Isomerisation:} The B-Z isomerisation of duplex DNA is a gross structural transition, from a right-handed double helix (the canonical B-form) to a left-handed double helix (the Z-form, which occurs {\it in-vitro} at high salt concentration \cite{Pohl1972,Ferreira2006} or {\it in-vivo} with the assistance of DNA-binding proteins \cite{Oh2002} and/or with negative supercoiling \cite{Peck1982}).  Occurrence of Z-DNA in mammalian cells has been implicated as a causative factor for certain cancers \cite{Wang2006}. In low salt the (more compact) Z-DNA is disfavoured relative to B-DNA for reasons of electrostatics and solvation; it becomes free-energetically favourable in total when a high salt concentration screens the repulsion between the charged backbone phosphate groups, also making partial solvent-exposure of the base pairs more favourable (\ref{fig:structs}).  The unwinding which is required to change the handedness of the double helix presents a formidable energetic barrier which must be overcome for the transition to take place \cite{Kastenholz2006,Lee2010}, even when the ends of the duplex are free.

The salt concentration for coexistence of these two forms is known from circular dichroism measurements to be roughly 2.8M NaCl for [d(CG)]$_6$; this varies with sequence, temperature, chain length and methylation \cite{Pohl1972,Behe1981,Ferreira2006}.  The availability of this experimental data, particularly for short dodecamer sequences, has attracted several simulation studies.  Definitive thermodynamic characterisation with respect to salt concentration has remained just over the horizon for some years, however recent improvements have been made: 3DRISM has been qualitatively successful in finding the coexistence point \cite{Maruyama2010} (estimated at 0.8M); and targeted molecular dynamics simulations have had at least some success in describing the transition path \cite{Kastenholz2006,Lee2010} although this last matter is still difficult to compare directly against experiment. 
 
{\bf Previous Calculations of the Energetics:} The history of computational studies of the B-Z transition from a free energy point of view is, in microcosm, the history of developing computational methods for molecular electrostatics and solvation.  Early treatments used continuum electrostatic models and rigid-body treatments of the DNA, either based on static atomic structures \cite{Kollman1982, Misra1996} or on geometric constructs such as grooved cylinders \cite{Gueron2000} or double-helical chains of charged beads \cite{Soumpasis1984}.  A review article in 1987\cite{Jovin1987} pointed out that results from continuum-electrostatics seemed to depend dramatically on the assumptions made for the various geometric and physical parameters.  The same review article also ruled out an all-atom simulation approach for the near future on grounds of computational expense (a view still advanced some 23 years later by researchers from another group\cite{Maruyama2010}).

A paper in 1997 \cite{Montoro1997} responded to the perceived limitations of a continuum treatment of the ions, by making a hybrid approach with discrete counterions in continuum solvent.  The same paper pointed out that treatments which give a coexistence concentration without calculating the non-electrostatic free energy difference are either right for the wrong reasons (if they ignore this contribution) or cheating somewhat, by setting the non-electrostatic free energy change to whatever value gives the correct coexistence point (e.g.~ref.~\cite{Gueron2000}).  In the `right for the wrong reasons' category falls one of the most accurate results, based on a classical density-functional theory (DFT) approach, which gave qualitative agreement of 3.6M NaCl to the infinite-chain coexistence value of 2.25M as early as 1989 \cite{Hirata1989}, however the model used was without any detailed representation of the DNA.  When the same author returned to the problem 21 years later with a similar but more detailed model taking into account the atomic structure of the water molecules and the DNA, the final value for the coexistence was in fact less accurate than the older one \cite{Maruyama2010}, at 0.8M (compared to the dodecamer coexistence value of ~2.8M), although of course more insight was available from the use of the richer model.

The calculation we present gives the most accurate value for the position of this conformational equilibrium from an all-atom treatment, and also the most accurate without fitting parameters to date (yielding a coexistence concentration of 2.5M NaCl compared to the 2.86M observed experimentally by Pohl \& Jovin in 1972\cite{Pohl1972}).  With the understanding that calculations from other methods can of course provide insight in different ways, the previous values reported in the literature were, in chronological order: 2.25M\cite{Soumpasis1984} (exactly right for the infinite-chain, but with Na$^+$ hard-sphere radius set as a free parameter to 4.95\AA~vs.~the crystal ionic radius of 0.12\AA\cite{Shannon1976}), 3.6M (vs.~2.25M for the infinite-chain)\cite{Hirata1989}, no transition\cite{Montoro1997}, 3.7M (vs.~2.4M given by them as the infinite-chain value)\cite{Gueron2000} and 0.8M\cite{Maruyama2010} (vs.~2.86 for the dodecamer).  

Given the difficulty of finding the correct coexistence value, some papers have drawn attention to the gradient of the free energy with respect to concentration, comparing their results with a fit to the empirical data of:

\begin{equation}
\beta \Delta G_{B-Z} = A \ln \frac{[{\rm Na}^+]}{C}  
\label{eqn:jofit}
\end{equation}

where $\beta=1/k_BT$, $\Delta G_{B-Z}$ is the free energy difference per base pair and $C$ is the coexistence concentration \cite{Pohl1983}.  This fit was derived based on data at the long-chain limit ($A=0.6$, $C=2.25$).  There has not been a great deal more success from theory in matching the observed gradient $\partial \beta\Delta G/\partial \ln [{\rm Na}^+]$ of the free energy near to coexistence than the coexistence point itself, and the fit of a simple logarithm to the free energy has been brought into question by more recent experimental work pointing to the existence of intermediates \cite{Ivanov1993,Grzeskowiak2005}. 

 The gradients near to coexistence of some reported theory treatments were:  $A=0.3$\cite{Misra1996}, $A=0.15$\cite{Montoro1997} and $A=0.3$\cite{Gueron2000}.  One theory treatment gave an elbow in the plot, with a value below coexistence of approximately $A=0.3$, and above coexistence of $A=0.6$\cite{Soumpasis1984}.

{\bf Thermodynamic Integration Relative to Exact Solution:} This established and powerful family of thermodynamic integration (TI) techniques has recently been extended to permit application to liquids \cite{Tyka2007,Schilling2009,Schmid2010}.   In brief, the method consists of defining for a given system a reference Hamiltonian $\mathcal{H}_1(\vec{r},\vec{p})$ which has an analytically tractable expression for its free energy (described in the supplementary data); and also a realistic but intractable $\mathcal{H}_0(\vec{r},\vec{p})$ (here, the AMBER Hamiltonian).  A third, `mixed' Hamiltonian is then defined as a sum of the two:

\begin{equation}
\mathcal{H}(\vec{r},\vec{p},\lambda) = f_0(\lambda)\mathcal{H}_0(\vec{r},\vec{p}) + f_1(\lambda)\mathcal{H}_1(\vec{r},\vec{p}).
\label{eqn:ham}
\end{equation}

Here $\lambda$ is introduced as a control variable of the system such that for $\lambda=0$, $f_0(\lambda)=1$, and $\mathcal{H}(\vec{r},\vec{p},0) = \mathcal{H}_0(\vec{r},\vec{p})$; while for $\lambda=1$,  $\mathcal{H}(\vec{r},\vec{p},1) = \mathcal{H}_1(\vec{r},\vec{p})$.   The Helmholtz free energy $A_0$ of the system under the realistic (but intractable) Hamiltonian $\mathcal{H}_0$ can then be expressed in terms of an integral with respect to $\lambda$:

\begin{equation}
A_0 = A_1 - \int_0^1  {\rm d \lambda} \,\left< \frac{\partial}{\partial \lambda}\mathcal{H}(\vec{r},\vec{p},\lambda)\right>_{N,V,T,\lambda}
\label{eqn:ti}
\end{equation}

where $A_1$ and $\mathcal{H}_1$ refer to the free energy and Hamiltonian of the reference system; $f_0()$ and $f_1()$ are mixing functions used to control the speed with which $\mathcal{H}_1$ is introduced in place of $\mathcal{H}_0$ over the interval $\lambda=[0,1]$; and $\langle \rangle_{N,V,T,\lambda}$ refers to the ensemble average for a given value of $\lambda$. Each `generalised force' $\left< \frac{\partial}{\partial \lambda}\mathcal{H}(\vec{r},\vec{p},\lambda) \right>_{N,V,T,\lambda}$ is found by collecting time-averages of the partial derivatives of $\mathcal{H}$ while the system evolves under this mixed Hamiltonian at fixed $\lambda$. (Here, obviously, the sampling times need to be long enough to allow for identification of the ensemble-average with the time-average.)  While the time-averaged generalised force may be slow to converge, this procedure has the advantage that calculations at each value of $\lambda$ can be carried out in parallel, without any need for intercommunication.

{\bf Einstein Molecule \& Cage-Swapping Liquid Model:} The first presentation of TIES was in the context of an $\mathcal{H}_1$ appropriate for solids or for macromolecules in vacuum, due to Frenkel and Ladd \cite{Frenkel1984} with modifications by Vega \cite{Vega2007}, and is known as the ``Einstein molecule method'' (EM) or as ``the confinement method''. The basic idea of the EM reference model is that the particles do not interact with each other, so that the reference partition function factorizes and the reference free energy can easily be computed exactly. In the simplest such model, the particles are simply coupled to predetermined reference positions by harmonic springs. 
 
The EM model was used here for the DNA chains. The reference model used for the water and counterions in this calculation, however, is still quite new,  and has been slightly modified from its most recent appearance in the literature \cite{Schmid2010} for use in conjunction with MD rather than Monte Carlo sampling.

\begin{figure}
\includegraphics[width=0.48\textwidth]{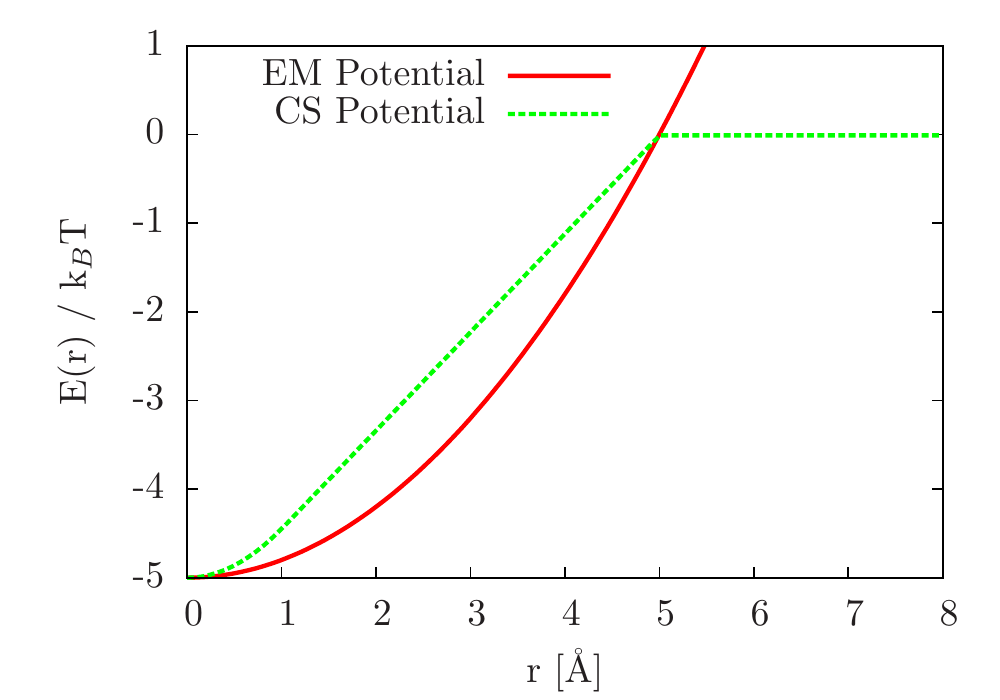}
\caption{A simple harmonic well was used for the EM potential, while a constant-force well with a cutoff at long range and a harmonic region at very short range was used for the CS potential of the fluid part of the system.}
\label{fig:emcs}
\end{figure}

The EM model of harmonic wells cannot be used unmodified as a reference for the liquid state. In the liquid each particle is ultimately free to move throughout the volume; so the range of the harmonic well in the EM model would have to be infinite as well. Thus to sample the average in (eq. 3) one would need to sample for infinite lengths of time. Instead we use a potential which is attractive at a short range from the reference position, but flat at long range (\ref{fig:emcs}). The contribution of the flat part to the partition function of the mixed model is trivial, so there is no need to extensively sample the reference model outside the short range of attraction.

The use of a potential cutoff solves the problem of dealing with an infinitely-ranged well of attraction, but it introduces another problem: for a large system volume, a diffusing particle hardly ever re-visits its reference site, once it has left the attractive well. We solve this problem by means of a Monte Carlo move in which we swap particle identities such that the diffusing particle has frequent opportunities (via a Metropolis acceptance rule) to rejoin or part from its reference site. This model is here called `CS', with reference to the `Cage-Swapping' dynamical behaviour which it provides.  The `Cage-Swapping' move reflects the factorial term in the general partition function of liquids, related to the indistinguishability of particles. For a detailed explanation see Schilling \& Schmid\cite{Schilling2009}.

The alternative in the literature to the CS MC move is to solve the linear assignment problem of particle identities so as to minimise the particle-well interaction energy at each MD timestep, as by Tyka, Sessions and Clarke \cite{Tyka2007} (see also \cite{Reinhard2007}).  This has both advantages and drawbacks over the approach of using an MC move, and a thorough comparison might be valuable in a later publication.  The principal advantage of linear assignment appears to be that the size of the (configuration $\times$ identity) space is reduced because the system always has the identity assignment which gives the mininum energy for the given configuration.  The principal advantage of MC cage-swapping would appear to be enhanced sampling, as nearby particles can exchange wells even if it is not energetically favourable to do so, with the well exchange creating an attractive force which pulls the particles to their new minima.

  Modifications made for the MD/MC CS model used here relative to previous pure-MC treatments were to use MD timesteps instead of those MC moves which moved particles spatially without changing their identities; and to add a harmonic region near to the well-bottoms; such that the wells were harmonic out to 1\AA, then constant-force out to 5\AA, then flat.  The small harmonic region was necessary for stability of the MD.  

A natural objection to the use of a swap move in conjunction with MD simulation is to ask if it is required that the MC sampling should converge before each MD timestep takes place; mercifully it has been shown in other work that any combination of MD and MC steps which would individually provide Boltzmann sampling of the degrees of freedom to which they are applied will in total provide Boltzmann sampling of the combined degrees of freedom \cite{LaBerge2000}.

{\bf Detailed Motivation of the Integration:} In a TI calculation, the free energy landscape of the system gradually morphs from one shape to another, in a way that is not yet easy to characterise.  From an intuitive basis, we can state qualitatively that for effective TI four sometimes-conflicting criteria are important:

\begin{enumerate}
\item[{\bf (a)}]{ To minimise the error from the numerical integration, the integration path should be as smooth as possible; meaning that the higher-order derivatives of each generalised force $\frac{\partial\,}{\partial \lambda}f(\lambda)\mathcal{H}$ should be as small as possible.}
\item[{\bf (b)}]{ The shapes of landscapes to be mixed should be similar to each other; otherwise the generalised force due to $\mathcal{H}_0$, for instance, could sporadically take extremely large values when $f_0(\lambda)$ is small and $\mathcal{H}_1$ is primarily controlling the dynamics.}
\item[{\bf (c)}]{ To minimise the convergence time per integration point,  exploration of the mixed landscapes should be as fast as possible.}
\item[{\bf (d)}]{ Because the variance of the generalised force scales with the gradient of the mixing function $f(\lambda)$, to efficiently distribute sampling between integration points, the mixing functions should be as close to linear as is consistent with the other criteria.}
\end{enumerate}

To meet criterion (a), it is first required to avoid discontinuities in the generalised forces by making sure that the integration does not cross any first order phase transitions.  The basis of TIES is a successful treatment of this issue by defining tractable models which match a given phase in terms of the symmetries expressed by their partition functions.  Definition of the model partition functions is discussed in the supplementary information.

Criterion (b) is more difficult. The reference model that we use decouples all particles from each other (in this way, the partition function is easily solved analytically). Thus at $\lambda=1$ particles can freely move through each other, while for $\lambda = 0$ this is not possible. Therefore the landscape under $\mathcal{H}_1$ is topologically different to that under $\mathcal{H}_0$. This brings the question of appropriately changing between landscapes sharply into focus.  In order to treat cases where a cavity in the free energy landscape (e.g.~a region of infinite potential energy, such as due to the Lennard-Jones short range repulsion) must be introduced, starting from a flat potential, it is valuable to gradually increase the maximum energy associated with that region to some high but finite value. The region is only then blocked completely, by introducing the infinite potential inside the region which is already effectively forbidden by the large finite potential. This strategy, which was attempted here by use of the `Guide Hamiltonian' discussed below,  should ensure that regions of phase space which are populated for a given $\lambda$ are also populated at neighbouring values of $\lambda$, while still allowing topological change of the landscape to occur.  In the absence of such a technique, the generalised force can diverge, with an infinitesimal change in $\lambda$ giving an infinite change in $\mathcal{H}\frac{\partial\,}{\partial \lambda}f(\lambda)$.

A further, partial, fix for problems with respect to criterion (b) is provided by choice of the mixing function used to introduce or remove a given Hamiltonian.  Careful work has been done on this in the context of alchemical TI involving Lennard-Jones and Coulomb forces \cite{Steinbrecher2007}, which was broadly followed here: the essence of the strategy is to use polynomial mixing functions such that as the ability of a given Hamiltonian $\mathcal{H}$ to direct the system into its own minima drops with $f(\lambda)$, the generalised force $\mathcal{H}\frac{\partial\,}{\partial \lambda}f(\lambda)$ also drops.  The choice of mixing function must be balanced between criteria (b) and (d).

Criterion (c) is addressed here by defining $\mathcal{H}_1$, the CS Hamiltonian, with few and low energetic barriers over the difficult ($\lambda <0.5$) part of the integration, such that exploration is relatively quick. It has been suggested that it  might also sometimes be valuable to retain barriers in the landscape  as these can serve to reduce the effective dimensionality of the configurational space and actually accelerate exploration \cite{McLeish2005}, however this effect was not observed here. 

Criterion (d) is addressed here by choosing a relatively low-order mixing function compared to some of those which have been trialled in other work \cite{Steinbrecher2007}.  

{\bf Guide Hamiltonian:} To manage the changes of the topology of the free energy landscape over the integration, a third Hamiltonian, $\mathcal{H}_{\rm guide}$, was introduced in addition to $\mathcal{H}_{0}$ and $\mathcal{H}_{1}$, with a mixing function such that it would be zero at both endpoints of the path.  The weak short-range interparticle repulsion provided by $\mathcal{H}_{\rm guide}$ controlled the topology in the sense that it was used to first introduce gently sloping energy barriers around those states which would later take on infinite energy becoming topological `cavities' in the free energy landscape.  The unshielded existence of such a cavity is intolerable if carrying out MD, because this leads to infinite values for the forces between particles.

This use of a guide Hamiltonian, introduced here, is an alternative to the practice of adding a $\lambda$-dependence to the functional form of the Lennard-Jones (LJ) potential such that it is finite over all configurations for $\lambda>0$, known as `softcoring' \cite{Beutler1994}. Softcoring of the LJ typically follows mixing-out of the Coulomb interaction (which also has a discontinuity at zero separation), in a multi-step procedure which increases computational expense. Such a process is also undesirable because it has the potential to introduce first order phase transitions: the phase diagram of water, for instance, is dramatically altered when electrostatics are removed.  Initial attempts were made to softcore the Coulomb and LJ terms together in such a way as to preserve the shape of the energy landscape and this was found to be prohibitively complex, especially given the requirement for the sake of efficiency to treat the Coulomb interactions using a fast algorithm such as the Ewald sum.

The use of an $\mathcal{H}_{\rm guide}$ has the advantages of solving at a stroke discontinuities of both LJ and Coulomb forces, of being extremely simple to implement and to parameterise, and of introducing no extra code into the complex and highly optimised non-bonded force and Ewald sum routines of the molecular dynamics program.  The short-range repulsion is handled using a  neighbour list and therefore adds only a small (and linearly scaling) cost to the calculation. 

The guide Hamiltonian was defined very simply as a quadratic repulsive potential between all heavy atoms, where the cutoff $R_{ij}$ was defined as the radius giving the minimum of the van der Waals interaction for the given atom pair $ij$, multiplied by 0.88. 

\begin{eqnarray}
\frac{\mathcal{H}_{\rm guide}}{k_BT} &=& 20\sum_{i=1}^N\sum_{j=i+1}^N \left(\frac{\mathrm{MIN}[ R_{ij} - r_{ij}, 0]}{R_{ij}}\right)^2
\end{eqnarray}

{\bf Mixing Function Design:} Numerical smoothness of the integration over $\lambda$ depends critically on the choice of mixing functions used to turn the Hamiltonians $\mathcal{H}_0$, $\mathcal{H}_1$  and $\mathcal{H}_{guide}$ off and on.  The mixing functions were defined as polynomials:

\begin{eqnarray}
f_0(\lambda)      &=& (1-\lambda)^4\\
f_1(\lambda)      &=& \lambda^2\\
f_{guide}(\lambda) &=& \frac{729}{16}\lambda^2(1-\lambda)^4
\end{eqnarray}

The choice of polynomial mixing functions (which has been advised with order 4 or greater for alchemical transformations involving Lennard-Jones atoms \cite{Steinbrecher2007})  gives generalised forces $\frac{\partial\,}{\partial \lambda} f_0(\lambda)\mathcal{H}_0$ and $\frac{\partial\,}{\partial \lambda}f_1(\lambda)\mathcal{H}_1$ which are zero for $\lambda=1$ and $\lambda=0$ respectively, so that (for example) two particles can pass through each other without causing a singularity in the generalised force with respect to $\mathcal{H}_0$.

Formally, the use of a polynomial mixing function is enough to stabilise the integration, mixing the two topologically distinct landscapes without introduction of singularities.  Unfortunately the multiplication of very large by very small numbers is numerically abhorrent, therefore the guide Hamiltonian was also required.  

\begin{figure}
\begin{center}
\includegraphics[width=0.4\textwidth]{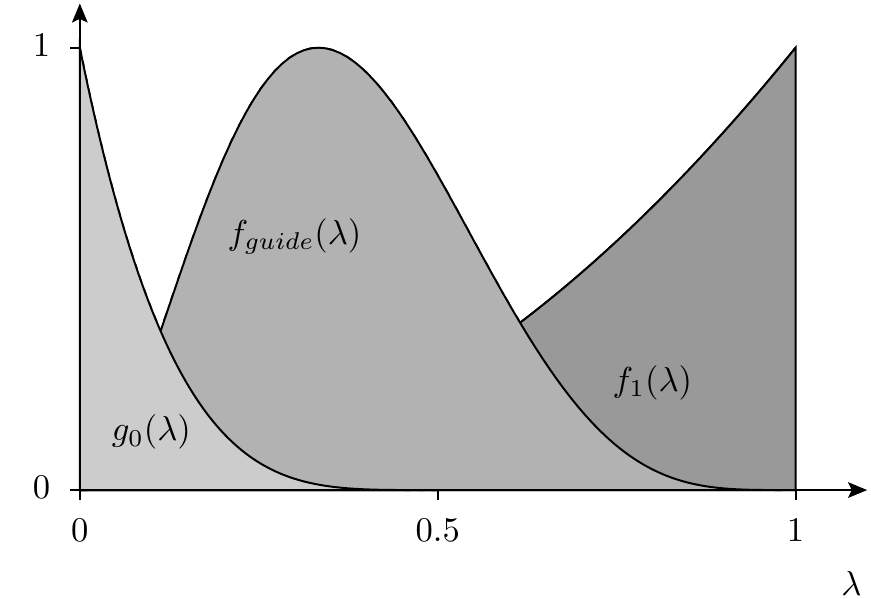}
\caption{The function $g_0(\lambda)$ controlled the AMBER Hamiltonian $\mathcal{H}_0$, $f_1(\lambda)$ controlled the analytically tractable reference Hamiltonian $\mathcal{H}_1$ and $f_{guide}(\lambda)$ controlled the `path' Hamiltonian $\mathcal{H}_{guide}$.}
\label{fig:mixFuncs}
\end{center}
\end{figure}

In order to allow $\mathcal{H}_{guide}$ to take effect before removing the non-smooth Hamiltonian $\mathcal{H}_0$, the mixing function $f_0()$, was rescaled such that the endpoint of the mixing would be at $\lambda = \lambda_g$ instead of at $\lambda=1$ (with $\lambda_g$ chosen arbitrarily as 0.5).  The actual function used in place of $f_0()$ was therefore: $g_0(\lambda)=\mathrm{MIN}[f_0(\frac{\lambda}{\lambda_g}), 0]$. The three mixing functions are shown in \ref{fig:mixFuncs}.

\section{Methods}
\label{sec:methods}

{\bf AMBER Setup:} Molar NaCl concentration was defined as the ratio of Na$^+$ ions to H$_2$O molecules multiplied by the molarity of water, 55.55.  $\mathcal{H}_0$ was defined using the AMBER99 forcefield \cite{Wang2000} with the Barcelona corrections to nucleic acid parameters \cite{Perez2007}, the Joung-Cheatham ion parameters \cite{Joung2008} and the TIP3P water model \cite{Jorgensen1983}.  Water molecules were kept rigid using the SHAKE algorithm.  Temperature was held at 300K using a Langevin thermostat with a coupling of 0.1ps$^{-1}$.  The seeds for the noise-generator in the Langevin thermostat were set to be the same for each pair of B and Z at a given $\lambda$ and [NaCl], in the hope of reaping some benefits by covariant sampling\cite{Assaraf2011} in the estimate of the free energy difference.  The molecular dynamics timestep was 2fs.  The `pmemd' dynamics engine\cite{Duke2003} provided in AMBER was used (with default parameters) to treat the electrostatic and other terms of the forcefield efficiently.

Start configurations of [d(CG)]$_6$ DNA double helices in the B and Z conformations were prepared using the NAB molecular building tool \cite{Macke1997}.  Counterions were added near to the DNA using the Xleap component of AMBER so as to neutralise backbone charge, and further counterions were then added at random to bring the salt concentration up to the required molarity assuming 9441 water molecules.  Water molecules were then added from a pre-equilibrated water box, deleting those which overlapped with an ion or DNA atom, until 9441 were present.  The concentrations studied were 1.11M, 1.61M, 2.11M and 2.61M NaCl.  The initial configurations were energy-minimised and then equilibrated at 300K and 1 atmosphere in the NPT ensemble for 2ns. The average volume of each system 1.11-2.61M NaCl was taken over the interval 1 to 2 ns, and the system box sizes in each configuration were set to these average volumes, so that the TI calculations could be carried out in the NVT rather than NPT ensemble.

Exploratory runs for the concentrations 1.11M to 2.61M NaCl were made out to 20ns to check stability and to estimate the important time and lengthscales of the dynamics (see supporting information).  Although the full TI calculations were not carried out at 2.86M NaCl, an initial 10ns test equilibration run was also made at this concentration (see supporting information). Molecular images were prepared using VMD\cite{Humphrey1996}.

{\bf Analytical Model Setup:} In the reference model of the system, all solute atoms (including hydrogens) were placed in harmonic wells of spring constant 5k$_B$T\AA$^{-2}$.   Counterions and water oxygen atoms were assigned to three sets of identical CS wells.  Because the TIP3P water molecules are treated as rigid triangles, it was not neccessary to restrain the water hydrogen atoms in addition to the oxygens.   The CS wells had a spring constant of  1.112k$_B$T\AA$^{-2}$\ up to $r_1=1$\AA, then a constant force of 1.11k$_B$T\AA$^{-1}$ out to $r_2=5$\AA. The `cage-swapping' reassignment of CS wells was carried out using a Metropolis MC algorithm.  For each liquid molecule present in the system, 5 MC well-swapping attempts were made every timestep. Each system run under a mixed Hamiltonian was allowed 500ps (250k timesteps) of equilibration, before collecting the generalised force time series for a minimum further 500ps, depending on convergence.  
\section{Results}
\label{sec:results}

The equilibrium work done on each integration path is shown in \ref{tab:freeEnergies}.

\begin{table}
\begin{center}
\begin{tabular}{ | l | c | c | }
\hline
    [NaCl]   &  $\Delta A_B$ ($k_BT$/bp)  & $\Delta A_Z$ ($k_BT$/bp)\\
\hline
     1.11M   &  3497.21 ($\pm$0.41)& 3495.60 ($\pm$0.40)\\  
     1.61M   &  4123.90 ($\pm$0.41)& 4123.20 ($\pm$0.40)\\
     2.11M   &  4751.90 ($\pm$0.42)& 4750.93 ($\pm$0.42)\\
     2.61M   &  5378.15 ($\pm$0.43)& 5378.35 ($\pm$0.44)\\
\hline
\end{tabular}     
\caption{Equilibrium work over each integration path, and estimated standard errors.}
\label{tab:freeEnergies}
\end{center}
\end{table}

\begin{figure}
\begin{center}
\includegraphics[width=0.49\textwidth]{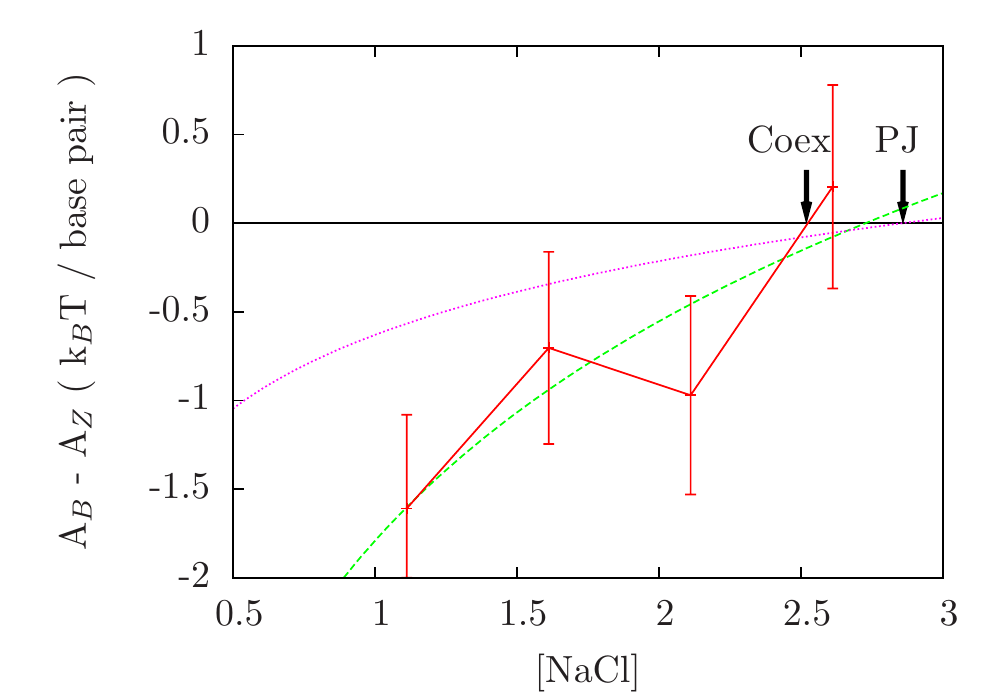}
\caption{Free energy differences between the B and Z conformations with respect to salt concentration.  The coexistence point found experimentally\cite{Pohl1972} is indicated with an arrow (PJ).  The dotted line is a fit by Pohl\cite{Pohl1983} to experimental data (eq.~1); the dashed line is a fit of the same function to our own data.}
\label{fig:freeEnergies}
\end{center}
\end{figure}

{\bf Free Energies:} Free energies were separately calculated for Z and B conformations over a range of salt concentrations (\ref{tab:freeEnergies} and \ref{fig:freeEnergies}).  The crossover between the B and Z regimes occurred at around 2.5M NaCl, quite near to the 2.86M found in  the experiments of Pohl and Jovin \cite{Pohl1972}. As far as the authors are aware, this is the most accurate numerical calculation of the coexistence position (without fitting free parameters) to date.  

The shape of the free energy difference away from coexistence does not seem to match the phenomenological equation proposed by Pohl\cite{Pohl1983}(eq.~1),  however given the small number of points and the large errorbars it is also difficult to completely discount this equation, or to suggest an alternative more complex functional form such as more recent theory \cite{Frank-Kamenetskii1985,Lukashin1991} and experiment\cite{Ivanov1993,Grzeskowiak2005} suggesting the existence of intermediates  seem to require.  A fit of eq.~1 with $A$ and $C$ free gives $A=1.8\pm0.7$ and the coexistence value $C=2.7\pm0.6$ M, although, because of the questionable agreement of the shape of eq.~1 with our data, we prefer to report coexistence as 2.5M based simply on the y-intercept of the trace between points three and four.

A possible explanation for the apparent overestimate of the free energy penalty for disfavoured forms (the large value of the fitting parameter $A$) relative to the experimental data, could be that the experimentally studied left-handed DNA structures at low salt are not the same as the conventional Z-DNA structure observed at high salt, which was used as the start configuration for the simulations at low salt.  If the starting configurations used were in fact strongly metastable with respect to some lower-energy left-handed form (as well as with respect to the still-lower B-DNA) then this would explain the observed behaviour.  A wider-ranging examination of the free energy landscape would be required to discuss this possibility further.

It was found that, while the Z-DNA structure used was strongly metastable at low salt concentrations, the more labile B-DNA structure tended to melt or fray at 2.86M NaCl and above (see supplementary data).  The greater flexibility of B-DNA, or the sharp elbow in the free energy difference above coexistence which has been suggested\cite{Soumpasis1984}, might serve to explain this.  Given that the Watson-Crick bonded B-DNA structure was unstable with respect to a disordered structure, the TIES calculation of this paper would have been difficult to apply without artificially enforcing W-C base pairing.  Separate to the difficulty of such a calculation, would be the question of its interpretation in a regime where it is no longer clear that B is the dominant right-handed metastable conformation.

\begin{figure}
\begin{center}
\includegraphics[width=0.49\textwidth]{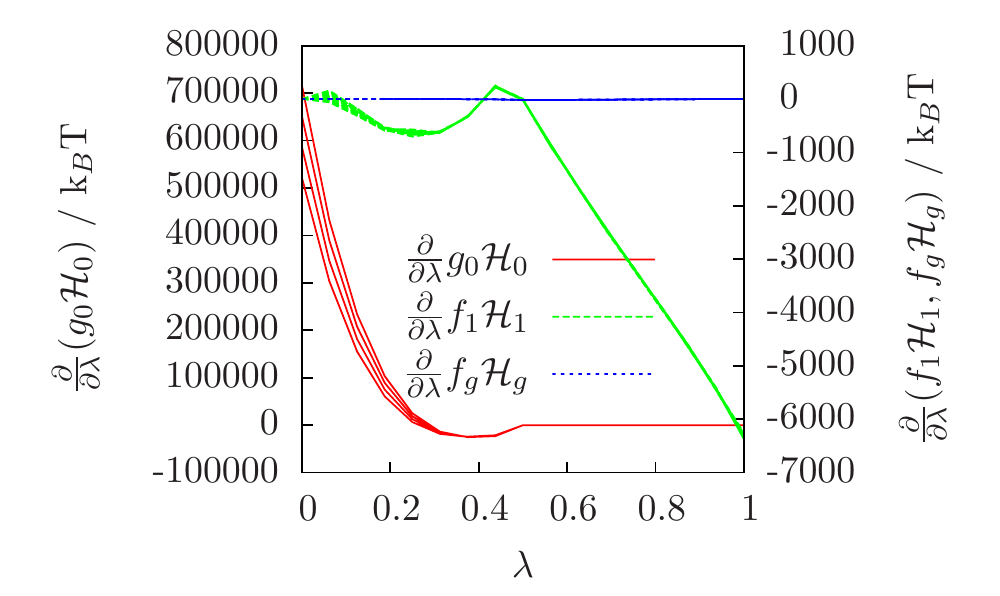}
\caption{Generalised forces due to the AMBER Hamiltonian ($\mathcal{H}_0$), the EM+CS Hamiltonian ($\mathcal{H}_1$) and the `guide' Hamiltonian used to prevent particle overlaps during the mid-section of the integration ($\mathcal{H}_{g}$). The guide Hamiltonian {\it (right axis)} was much smaller than the main Hamiltonians {\it (left \& right axes)}. A trace is shown for each of the eight calculations carried out, however B and Z cannot be distinguished on the scales used.}
\label{fig:genForce}
\end{center}
\end{figure}

{\bf Integration Path:} The character of the integration path can be seen by examining the generalised force due to each of the three Hamiltonians with respect to $\lambda$ (\ref{fig:genForce}).  The smoothness of these curves (apart from at $\lambda=0.5$, where the integration-out of the AMBER Hamiltonian is completed) indicates that no first-order phase transitions were crossed during the integration path, and also serves to justify the use of a basic Simpson's rule integration scheme; carried out separately over the three terms of the generalised force and over the [0,0.5] and the [0.5,1.0] intervals.

\section{Performance and Convergence}
\label{sec:covar}
Calculations were carried out using four 2.26 GHz processor cores per integration point (therefore 68 per measurement of $A_{B}-A_{Z}$).  Calculations required on average 26.3 hours per 100ps block giving a total of 210,000 core-hours for the entire calculation (42 core-months assuming 24/7 operation), plus equilibration.  

\begin{figure}
\begin{center}
\includegraphics[width=0.49\textwidth]{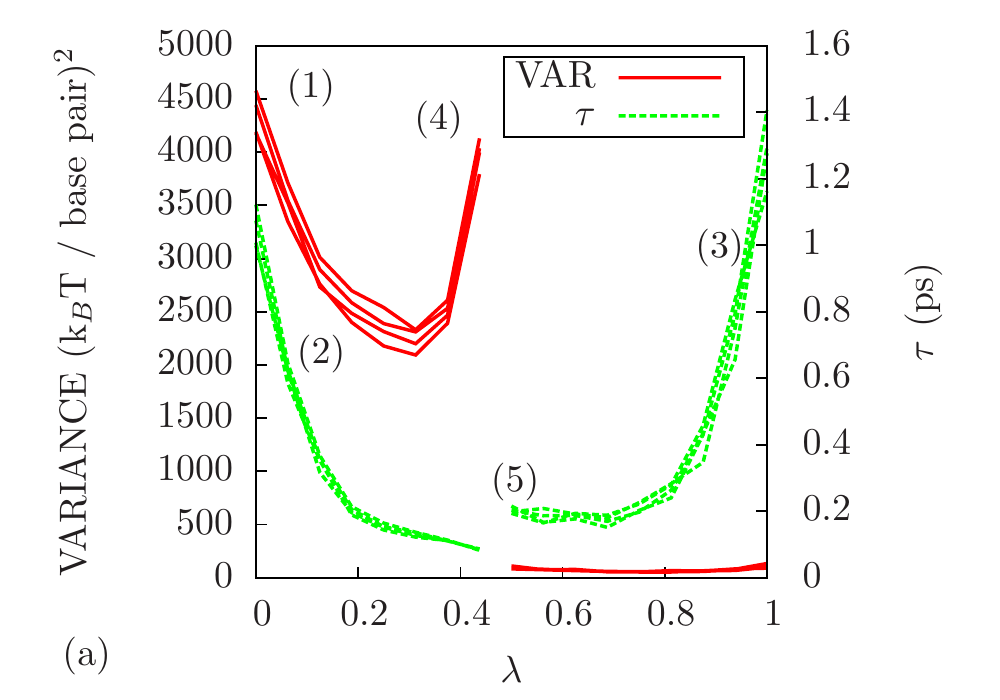}
\includegraphics[width=0.49\textwidth]{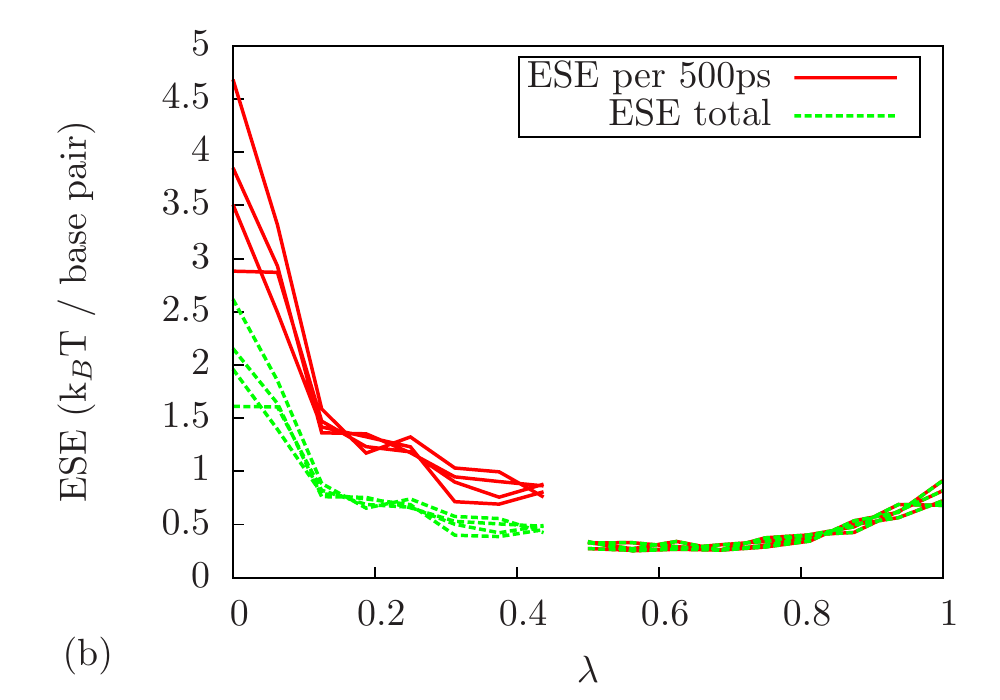}
\caption{{\bf (a) Variance and Decorrelation Time of Generalised Force}.  The rate of convergence of the estimate of the summed generalised force (B-Z) at a given $\lambda$ was determined by variance (left axis, red) and by the decorrelation time $\tau$ (right axis, green). Numbered labels indicate peaks in the variance or in $\tau$ which are discussed individually in the text. {\bf (b) ESE of the Generalised Force}. The ESE per 500ps was much higher for $\lambda < 0.5$,  therefore longer runs of 1600ps were made for this section.  A separate trace is shown for each salt concentration studied.  Lines are not drawn between the points at $\lambda=7/16$ and $\lambda=8/16$ because the generalised forces on the two intervals $\lambda<0.5$ and $\lambda\ge0.5$ had different shapes and were integrated separately. Each trace corresponds to the $\Delta A$ calculation for a given [NaCl].}
\label{fig:covar}
\end{center}
\end{figure}

To estimate the convergence of the individual generalised force measurements at each integration point, a spectral analysis of the time series of generalised forces was carried out using the CODA library of R functions \cite{Plummer2006}, in order to identify the decorrelation times and effective numbers of independent samples present.  Estimated Standard Errors of the mean (ESE) are given as the root of the variance divided by the effective number of independent samples. The ESE at each value of $\lambda$ (\ref{fig:covar}(b)) provides a description of the efficiency of the sampling in each of the `mixed' free energy landscapes which were visited.  

The radically different sampling efficiencies at the different integration points derive from separately addressable causes, the signatures of which are individually labelled in \ref{fig:covar}, and which are discussed in relation to the criteria for good TI design which were presented above:

\begin{enumerate}
\item[{\bf (a)}] {{\bf Smoothness of Generalised Force w.r.t. $\lambda$:} The smoothness criterion was met quite well (\ref{fig:genForce}).}

\item[{\bf (b)}] {{\bf Small Variance of Generalised Force:} When the potential corresponding to a given term of the generalised force is very different to the potential which is controlling the dynamics, then fluctuations in that term of the force can become very large, the peak labelled `4' on \ref{fig:covar} shows this effect.

Peak `4' is lower than in first attempts, thanks to the use of a polynomial mixing function, and also greatly diminished by the use of the guide Hamiltonian.  Peak `4' might also be further reduced by the use of different $\mathcal{H}_1$ Hamiltonians which more closely mirror the dynamics of the `real' $\mathcal{H}_0$ system.}

\item[{\bf (c)}] {{\bf Fast Exploration:} The peaks labelled `2' and `3' show slow exploration due to roughness of the potential energy landscape: sampling of the Boltzmann distribution for atomistic models of biomolecules is famously difficult, due to the many traps and barriers which exist.  Sampling also becomes slower in the CS Hamiltonian when it is fully turned on, due to the depth of the wells becoming greater than the thermal energy.   Peak `5' shows a slowing of exploration due to decoupling of particles under the CS Hamiltonian: when the particles no longer interact, decorrelation over time must rely purely on the Langevin thermostat rather than being assisted by the chaotic properties of the many-body Hamiltonians.  Although the aspect indicated by peak `5' seems to be of relatively minor importance, it should be possible to address it by altering the coupling parameter of the Langevin thermostat for different values of $\lambda$, or by using an $\mathcal{H}_1$ which has explicit interparticle forces while remaining tractable.

To address point (c) there is no easy escape from the need to sample the atomistic model at one end of the integration (`2').  At the other end of the integration (`3'), the trapping due to the CS potential could (and probably should) be reduced in a future calculation simply by holding the CS potential constant for $\lambda>0.5$.}

\item[{\bf (d)}] {{\bf Near-linear Mixing:} Peak `1' arises because the gradient of the mixing function $g(\lambda)$ is large at small $\lambda$, and this large value scales not only the generalised force but its variance - this represents a tradeoff accepted in the process of mixing function selection, it stands in balance against peak `4'.}
\end{enumerate}

To address this need to trade between the criteria (b) and (d) it seems that the best hope is to improve the guide or endpoint Hamiltonians used such that the need for a polynomial mixing function is removed.

\section{Discussion}

This calculation is the most quantitatively accurate numerical estimate of the salt concentration at B-Z DNA coexistence to date.  As recently as 2010 it was noted that an atomistic calculation of this value should be prohibitively difficult ``because the problem requires the sampling of an extraordinarily large configuration space, including water and ions, to obtain the free energy difference'' \cite{Maruyama2010}; that the result is quantitatively correct demonstrates solid improvements in the techniques of free energy estimation suitable for biomolecular complexes, over and above the normal progress due to improved hardware. 

 Because of the reductive nature of the molecular dynamics approach (treating, for instance, an atom in a protein in much the same way as an atom in a DNA base) it is reasonable to assume that the software and parameter sets developed can be reused widely and can be effective for any combination of protein, nucleic acid, counterions and solvent; although there is still substantial room for further development of them. It is intended to release the software and parameter sets as a downloadable library on Dr Berryman's website, for easy linking against the AMBER simulation program or other packages.   

The accuracy of the final result for the coexistence serves as a validation for use of the combination of the Joung-Cheatham ion parameters and the parmbsc forcefield (both of which are relatively new) with TIP3P water for simulation of DNA in the fairly unusual conditions of very high salt concentration.  

The work here has made minor alterations to the `Cage-Swapping' model for absolute free energy calculation of fluids and combined it with a standard Einstein Molecule method for the solute.  The principal novelty was in the scale and complexity of the system treated, but the introduction of a `guide' Hamiltonian to control the path of the thermodynamic integration while not altering behaviour at the endpoints is also novel as far as the authors are aware.  

There is too much active research in thermodynamic integration and free-energy perturbation methods to give a comprehensive list of ideas which could influence further work. A brief suggestion is that absolute free energy methods could be well suited to mapping of phase diagrams of complex fluids.  There are a large number of biomolecular and soft-matter systems which exhibit rich phase behaviour that have not yet been explored.  On the methodological side, recent advances in nonequilibrium TI \cite{Vaikuntanathan2011}, in correlated sampling \cite{Assaraf2011} and in Hamiltonian-exchange \cite{Khavrutskii2010} all offer increases in computational efficiency for future TIES calculations. It has been suggested that an efficiency gain can be made by doing away with the intermediate stages of the TI calculation entirely, instead carrying out a type of importance sampling over $\mathcal{H}_0$ and $\mathcal{H}_1$ such that those areas of configurational space which overlap between them receive enhanced attention \cite{Wilms2012}.

If one is interested in locating a phase transition one can, of course, use eq.~3 to compute the free energy difference between two phases directly and avoid the ``detour'' via absolute free energies; however, eq.~3 only holds if the integration path does not cross a first order phase transition. TIES is thus more generally useful (compared to performing TI over single or multi-legged free energy cycles between non-tractable Hamiltonians) in that states which are separated by a first order phase transition can be compared.  This is much simpler than carrying out TI calculations which integrate between phases by following a complex path designed to give consistently pseudo-critical behaviour \cite{Grochola2004,Eike2006}. 

A second general argument for TIES, which is more germane for the specific calulation presented here, is that setting one endpoint of the integration as an analytically tractable model leaves open the possibility of a shorter and smoother integration path, with smaller variances in the generalised force at a given $\lambda$, as the analytical models are refined. The most obvious step in further development, then, is extension and refinement of the analytical models which serve as endpoints for the calculation.  Ideally the mixed energy landscapes generated by an analytically tractable model should be as similar as possible to `softened' versions of the initial Hamiltonian.  This possibility should serve as a call to arms for any theorists who may believe that they can swiftly assemble a tractable model for some system of interest.

A supporting document is provided, giving  derivations for the absolute free energies of the CS and EM Hamiltonians. A second supporting document showing convergence of the simulations is also provided. This material is available free of charge via the Internet at \url{http://pubs.acs.org}. 

\acknowledgement
This work was greatly assisted by discussions with Charles Laughton and Sarah Anne Harris. Computing was provided by the HPC facility of the University of Luxembourg and by the European Soft Matter Infrastructure program.


\providecommand*{\mcitethebibliography}{\thebibliography}
\csname @ifundefined\endcsname{endmcitethebibliography}
{\let\endmcitethebibliography\endthebibliography}{}

\end{document}